\documentclass[%
 reprint,
 amsmath,amssymb,
 aip,
 jmp,
]{revtex4-1}
\usepackage[pdftex]{graphicx}
\usepackage{dcolumn}
\usepackage{bm}
\usepackage[colorlinks=true,linkcolor=blue]{hyperref}

\begin{document}

\preprint{APS/123-QED}

\title[Wide-range shell correction to the Thomas--Fermi theory and equation of state for electrons]{Wide-range shell correction to the Thomas--Fermi theory and equation of state for electrons}

\author{Sergey Dyachkov}
 \email{serj.dyachkov@gmail.com}
\affiliation{
Joint Institute for High Temperatures of the Russian Academy of Sciences
}%
\affiliation{%
Moscow Institute of Physics and
Technology, Institutskii per. 9, Dolgoprudny, Moscow Region, 141700, Russia}
\author{Pavel Levashov}%
 \email{pasha@ihed.ras.ru}
\affiliation{
Joint Institute for High Temperatures of the Russian Academy of Sciences
}%
\affiliation{%
Tomsk State University, 36 Lenin Prospekt, Tomsk, 634050, Russia
}%

\date{\today}
\begin{abstract}
Shell effects reflects irregularities of physical quantities caused by a discrete energy spectrum. The theory of the shell effects by Kirzhnits and Shpatakovskaya is valid only at relatively low densities providing for oscillations of thermodynamic functions. Similar effects for the electronic binding energy of a neutral atom were considered by Englert and Schwinger. In this work we propose a method of calculation of shell effects applicable in a wide range of density and temperature. The model is based on the finite-temperature Thomas-Fermi theory. Shell corrections to thermodynamic functions are obtained by special accounting of semiclassical states of bound electrons in the Thomas-Fermi potential. The results are in good correspondence with the precise Saha approach for the low density plasma and density functional theory simulation at high density. 
\begin{description}
\item[PACS numbers]
52.27.Gr, 52.25.Kn
\end{description}
\end{abstract}



\maketitle
\section{\label{sec:intro} Introduction}

Wide range equations of state (EOSs) are of practical importance in various applications of physics of extreme states of matter. The high quality EOSs may be constructed nowadays in computer simulations of many-particle quantum systems. The state-of-the-art first principle approaches such as density functional theory (DFT) overcome existing statistical theories. Unfortunately, the complexity of such calculations at temperatures about tens of $eV$ and higher becomes the major restriction for modern computers. Still, it is necessary to obtain the properties of warm dense matter at these conditions as, for example, for problems of interaction of intense energy fluxes with matter. 

The latest improvements in the average atom models seem to have enough precision to cover the gap of strongly coupled plasma in wide range EOSs between DFT and ideal gas. Moreover, such approaches do not require much computational resources due to deep theoretical analysis and effective simplifications. The Thomas-Fermi model was first developed for a single separate atom \cite{Thomas:1927, Fermi:1927} and had evident shortcomings. The idea was to consider an electron cloud surrounding the nucleus as a many-particle statistical system and interacting within a self-consistent electrostatic field. This semiclassical approach was good enough for heavy atoms with a large $Z$ and became the basis for the further development of DFT. But it turned out that  bound electrons near the nucleus were inadequately reproduced with infinite density. Exchange effects were also out reach of this theory what led to the improper dependency for the total binding energy.

Further the model was essentialy modified by various authors. Dirac \cite{Dirac:1930} came with an exchange correction which was introduced as an exchange energy in a density functional. Weizsacker \cite{Weizsacker:1935} figured out the result of rapid change of the electron density near the nucleus as gradient correction to the kinetic energy. Scott \cite{Scott:1952} introduced the correction to the total binding energy by excluding improperly accounted electron states in the Thomas-Fermi atom.

But for practical applications the finite-temperature model was also required \cite{Feynman:PR:1949}. This approach was good enough at high pressures and temperatures but still there were problems described above. Kirzhnits \cite{Kirzhnits:JETP:1957} introduced the generalized consistent method for accounting for quantum and exchange effects at finite temperatures. The density matrix was expanded to the required order of a semiclassical parameter in a basis of plane waves and necessary corrections appeared in the second order. Later, Kalitkin \cite{Kalitkin:IAM:1975} calculated detailed tables of thermodynamic functions for the finite-temperature Thomas-Fermi model with quantum and exchange corrections. This outstanding result claimed to be a solution for the problem of wide-range equations of state but still shell effects were unaccounted.

The recovery of shell effects were done in the paper \cite{Zink:PR:1968} in the case of high pressure ionization. In the major review of the semiclassical model \cite{Kirzhnits:SovPhysUsp:1975} the shell structure of atoms had been also discussed, but the solution had not been found in the general case. All the effects may be reproduced with an exact solution of the Schr\"odinger equation for electrons as it is done in the Hartree-Fock-Slater approach \cite{Nikiforov:2000}, but in the Thomas-Fermi model the following reduction of the quantum effects takes place \cite{Shpatakovskaya:SovPhysUsp:2012}:
\begin{multline}
  \rho(r) = 2\sum_{\varepsilon_n \leq \mu}|\psi_n(r)|^2 \mathrm{(gradient\, effects)} \sim \\
  \sim 2\sum_{\varepsilon_n \leq \mu}|\psi^{osc}_n(r)|^2 \mathrm{(oscillatory\, part)} \sim \\
  \sim 2\sum_{\varepsilon_n \leq \mu}\frac{|c_n|^2}{p_n(r)} \mathrm{(shell\, effects)} \sim \\
  \sim 2\int_{\varepsilon_n \leq \mu}\frac{|c_n|^2}{p_n(r)}dn \sim \rho_{TF}(r).
\end{multline}
The semiclassical approach is good enough for the continuous spectrum and, indeed, the main difference from the exact approach to the average atom is in the bounded states of electrons. This difference may be recovered as a correction to Thomas-Fermi model. 

The importance of shell effects for high quality EOSs at finite temperatures was also discussed by Iosilevsky and Gryaznov \cite{Iosilevsky:TVT:1981} in terms of comparison of the low-pressure isobars of Li and Na calculated by the Thomas-Fermi model with quantum and exchange corrections \cite{Kalitkin:IAM:1975} and the chemical picture model of plasma (known also as Saha model). The ionization of electron shells with heating leads to a stepwise increase in pressure and energy of the system what can be clearly observed in the Saha model with experimental ionization potentials. The Thomas-Fermi model with the existing corrections was unable to reproduce such effect without accounting for the bound electron states. 

The shell corrections at zero temperature were calculated precisely and appeared to have strong enough effect on atomic binding energy \cite{Englert:PRA:1985}. For practical needs the finite--temperature model was developed by Galina Shpatakovskaya. In the earlier works \cite{Shpatakovskaya:TVT:1985} the shell correction was calculated only in the case of high temperatures and low densities. The results were compared with the Saha model and here the great achievement was reached: the specific oscillations of thermodynamic functions appeared. The next step was to extend the method into the region of high densities and lower temperatures. A lot of modifications were done, even accounting of the zone structure. It was shown \cite{Shpatakovskaya:1998} that the method is applicable up to the normal density and further. The significant analytical analysis allowed to reduce the complexity of calculations so that the model remains as simple as the Thomas-Fermi approach. Moreover, it was developed in a self-similar manner, so that the results for hydrogen may be transformed for any element. It seemed that the problem of shell effects in the Thomas-Fermi model was solved.

Unfortunately, the authors of this work encountered some problems while implementing these methods for the wide-range EOSs. It was found out that the shell correction was derived in different approximations for the low and high temperatures. These approaches had to be joined between 10 and 100 $eV$ and there was no universal method to do it accurately for any element. It was inconvenient to use such a model for the wide--range calculations, so that the alternative approaches for accounting the shell effects in a more stable and precise way were developed \cite{FTTFQES}. The obtained technique requires more calculations, but remains computationally effective for modern computers. It covers the whole range of elements for the temperatures and densities in the region of validity of the Thomas-Fermi model with quantum and exchange corrections \cite{Dyachkov:PhysPlasmas:2014}.

\section{\label{sec:FTTF} Semiclassical thermodynamics of the Thomas-Fermi atom}
All further equations are written using the atomic system of units ($ e = 1 $, $ \hbar = 1 $, $ m_e = 1$).

\subsection{\label{subsec:TF_finite_T_general} Free energy of many electron system}

Free energy of a system consisting of $N$ electrons in the volume $V$ and at the temperature $T$ is the sum over quantum states $\{n\}$ with the electron-electron interaction energy in the self-consistent field $U_{e}(\mathbf r)$:
\begin{multline}
  \label{subsec:FTTF:eq:free_energy}
  F(V, T, N) = 
    -T\sum_{\{n\}} \ln
    \left[
        1 + \exp\left(\frac{\mu - \varepsilon_{n}}{T}\right)
    \right] + \\
    + \frac12 \int \rho_e({\mathbf r})U_{e}({\mathbf r})d^3r  + \mu N.
\end{multline}
Here $\mu$ -- the chemical potential and $\rho_e$ -- the electron density. If the Hamiltonian $\hat H$ is defined for that system, one can find that
\begin{equation}
  \sum_{\{n\}} \ln
    \left[
        1 + \exp\left(\frac{\mu - \varepsilon_{n}}{T}\right)
    \right] 
    = \mathrm{Tr}\,\ln
    \left[
        1 + \exp\left(\frac{\mu - \hat H}{T}\right)
    \right].
\end{equation}

The electron potential can be written as:
\begin{equation}
  U_{e}(\mathbf r) = -\int \frac{\rho_e(\mathbf r')d^3 r'}{|\mathbf r - \mathbf r'|}.
\end{equation}
If the system of electrons is considered with ions, then the total potential is:
\begin{equation}
  U(\mathbf r) = U_e(\mathbf r) + U_i(\mathbf r),
\end{equation}
where
\begin{equation}
  U_i(\mathbf r) = \sum_k \frac{Z_k}{|\mathbf r - \mathbf r_k|}.
\end{equation}
The self-consistent potential for a system of electrons and ions satisfies the Poisson equation:
\begin{equation}
  \label{FTTF_finite_T:eq:poisson_general}
  \Delta U (\mathbf r) = 4\pi \rho_e(\mathbf r) - 4\pi \sum_k Z_k\delta(\mathbf r - \mathbf r_k)
\end{equation}

The electron-electron and electron-ion interaction energies
\begin{equation}
  E_{ee} = -\frac12 \int \rho_e(\mathbf r)U_{e}(\mathbf r)d^3,
\end{equation}
\begin{equation}
  E_{ei} = -\int \rho_e(\mathbf r)U_{i}(\mathbf r)d^3r,
\end{equation}
along with the condition
\begin{equation}
  \label{subsec:FTTF:eq:number_of_states}
  N = \int \rho_e(r) d^3r,
\end{equation}
lead to the following expression for the free energy:
\begin{multline}
  \label{subsec:FTTF:eq:free_energy2}
  F(V, T, N) = 
    -T\,\mathrm{Tr}\,\ln
    \left[
        1 + \exp\left(\frac{\mu - \hat H}{T}\right)
    \right] + \\
    + E_{ee} + E_{ei} + \frac12 \int \rho_e({\mathbf r})(\mu + U({\mathbf r}))d^3r.
\end{multline}

If the free energy of the system is known one can proceed to calculations of thermodynamical properties. The first derivatives:
\begin{equation}
  \begin{aligned}
    & \mathrm{pressure:} & P = -F'_V, \\
    & \mathrm{entropy:}  & S = -F'_T, \\
    & \mathrm{energy:}   & E = F - TF'_T.
  \end{aligned}
\end{equation}
The second derivatives:
\begin{equation}
  \begin{aligned}
    & \mathrm{heat\,capacity:}  & C_V = -TF''_{TT}, \\
    &                           & C_P = -TF''_{TT} + \frac{T(F''_{VT})^2}{F''_{TT}}, \\
    & \mathrm{sound\,speed:} & c^2_T = V^2F''_{VV}, \\
    &                           & c^2_S = V^2F''_{VV} - \frac{V^2(F''_{VT})^2}{F''_{TT}}.
  \end{aligned}
\end{equation}

In general it is quite difficult to obtain all these properties: one have to calculate electron density in the self-consistent field and then find all the occupied states. Therefore, we begin with a simple approach based on the Thomas-Fermi approximation.

\subsection{\label{subsec:TF_finite_T_general} Thomas-Fermi approximation}

To calculate the trace in the free energy expression \eqref{subsec:FTTF:eq:free_energy2} let us consider it in a planewave basis $ \psi_{s\sigma}(\mathbf r, \mathbf p) = (2\pi)^{-3/2}e^{i\mathbf p \mathbf r}\delta_{s \sigma} $. For some operator $ \hat O $ we get:
\begin{equation}
  \label{FTTF_finite_T:eq:trace_operator}
  \mathrm{Tr}\,\hat O = \sum_{s = \sigma}\int d^3r\, d^3p\, \psi^{*}_{s\sigma}(\mathbf r, \mathbf p)\hat O \psi_{s\sigma}(\mathbf r, \mathbf p)
\end{equation}

First, we consider the hamiltonian for a system of electrons that includes only the kinetic energy and Coulomb interaction:
\begin{equation}
  \label{FTTF_finite_T:eq:hamiltonian}
  \hat H = \frac{\hat p^2}{2} - U(\mathbf r), \quad \hat p = -i \nabla.
\end{equation}
To avoid exponents in the integral \eqref{FTTF_finite_T:eq:trace_operator} one can replace momentum operator with $\hat p = \mathbf p - i \nabla $. Thus, hamiltonian here consists of two noncommutative summands. To calculate the function of them we use the expansion:
\begin{multline}
  \label{FTTF_finite_T:eq:commutator}
  f(\hat a + \hat b) = f(a + b) 
  + \frac12[\hat a, \hat b]f''(a + b) + \\
  +\frac16\left([\hat a,[\hat a, \hat b]] 
  + [[\hat a, \hat b],\hat b]\right)f'''(a + b) + \\
  + \frac18([\hat a, \hat b])^2f^{IV}(a + b) + ...
\end{multline}

In the Thomas-Fermi approximation only a zero term is accounted (higher order terms will be accounted further as corrections), thus the trace:
\begin{multline}
  \label{FTTF_finite_T:eq:free_energy_trace}
  F_{\mathrm{Tr} } = -2T\iint \frac{d^3rd^3p}{(2\pi)^3}
      \ln\left[1 + \exp\left(\frac{\mu - U(\mathbf r) - p^2/2}{T}\right)\right] = \\
      = -\frac{\sqrt2}{\pi^2}T^{5/2}\int \frac23 I_{3/2}\left(\frac{\mu + U(\mathbf r)}{T}\right)d^3r
\end{multline}
where the momentum integral is calculated by parts with the change $ \varepsilon = p^2/2$. By setting $\Phi(\mathbf r) = (\mu + U(\mathbf r))/T $ we get the expression for the free energy in the Thomas-Fermi approximation:
\begin{multline}
  \label{FTTF_finite_T:eq:free_energy}
  F_{TF} = \frac{\sqrt2}{\pi^2}T^{5/2}\int
           \left[\Phi(\mathbf r)I_{1/2}(\Phi(\mathbf r))\right. -  \\ - \left.\frac23 I_{3/2}(\Phi(\mathbf r))
           \right]d^3r
           + E_{ei} + E_{ee}.
\end{multline}

The only unknown quantity in the free energy expression is the potential $ U(\mathbf r) $. There are several possibilities to evaluate it. For example, one can use density functional theory approach and evaluate electron density at the free energy minimum. In this work we will use spherical cell approximation to evaluate electron density for a single average atom.

\subsection{\label{subsec:FTTF_finite_T_potential} Thomas-Fermi potential}

To solve the Poisson equation \eqref{FTTF_finite_T:eq:poisson_general} explicitly one have to define the electron density. It can be expressed using the density matrix operator:
\begin{equation}
  \label{FTTF_finite_T:eq:dens_matrix}
  \hat \rho_e (\hat H) = \frac{1}{1 + \mathrm{exp}\left(\frac{\hat H - \mu}{T}\right)},
\end{equation}
Similarly to free energy evaluation, it can be calculated using a planewave basis:
\begin{equation}
  \label{FTTF_finite_T:eq:denisty}
  \rho_e(\mathbf r) = \sum_{s = \sigma}\int d^3p\, \psi^{*}_{s\sigma}(\mathbf r, \mathbf p)\hat \rho(\hat H) \psi_{s\sigma}(\mathbf r, \mathbf p)
\end{equation}
Zero order of the density matrix expansion corresponds the Thomas-Fermi approximation, and the electron density proceeds to:
\begin{equation}
  \label{FTTF_finite_T:eq:density_TF}
  \rho_e(\mathbf r) = 
    \int
    \frac{1}{1 + \exp\left(\frac{p^2/2 - U(\mathbf r) - \mu}{T}\right)}
    \frac{2d^3p}{(2\pi)^3}.
\end{equation}
Thus, by solving the poisson equation one can obtain the self-consistent field and the electron density. 

From here we will use a single atom approach with a spherical symmetry of a self-consistent field $U(r)$. The Thomas-Fermi density of electrons $\rho_{TF}$ then:
\begin{multline}
  \label{FTTF_finite_T:eq:density_TF_spheric}
  \rho_{TF}(r) = 
    \frac{2}{(2\pi)^3}\int_0^{\infty}
    \frac{4\pi p^2\,dp}
    {1 + \exp\left(\frac{p^2/2 - U(r) - \mu}{T}\right)} = \\
    = \frac{(2T)^{3/2}}{2\pi^2}I_{1/2}\left(\frac{U(r) + \mu}{T}\right).
\end{multline}

The boundary problem for the potential is considered within a spherical cell of the radius $ r_0 $ and the volume $V = 4\pi r_0^3/3$. As the potential can be defined with an arbitrary constant value it is convenient to set $U(r_0) = 0$. The electroneutrality of an atomic cell lead to zero electric field outside (Gauss theorem) thus $U'(r_0) = 0$. Near the nucleus the potential should be coulomb, thus we have $rU(r)|_{r \rightarrow 0} = Z$ and the boundary problem:
\begin{equation}
  \label{FTTF_finite_T:eq:boundary_problem}
  \left\{
    \begin{aligned}
      \frac{1}{r}\frac{d^2}{dr^2}(rU)
      &=
      \frac{2}{\pi}(2T)^{3/2}I_{1/2}
      \left(
        \frac{U(r) + \mu}{T}
      \right), \\
      rU(r)|_{r=0} &=Z,
      U(r_{0})=0,
      \left.
        \frac{dU(r)}{dr}
      \right|_{r=r_{0}}=0. \\
    \end{aligned}
  \right.
\end{equation}

This boundary problem can be reduced to the dimensionless one with the change of variables $ \phi(x)/x = U(r) + \mu $, $ xr_0 = r $,  $a=4\sqrt{2}r^2_0/\pi $:
\begin{equation}
  \label{FTTF_finite_T:eq:boundary_problem_reduced}
  \left\{
    \begin{aligned}
      \frac{d^2\phi}{dx^2} &=
      axT^{3/2}I_{1/2}
      \left(
        \frac{\phi(x)}{Tx}
      \right), \\
      \phi(0) &=\frac{Z}{r_0},
      \quad \phi'(1)=\phi(1)=\mu.
    \end{aligned}
  \right.
\end{equation}
The function $\phi$ changes fast at small values of $x$, and for practical calculations it is better to use variable $u = \sqrt x$. The functions for integration can be defined as follows: $W = \phi - u^2\mu$, $2uV = W'_{u}$. Then boundary problem is:
\begin{equation}
  \label{FTTF_finite_T:eq:boundary_problem_reduced2}
  \left\{
    \begin{aligned}
      & V'_u = 2au^3T^{3/2}I_{1/2}\left(\frac{W + u^2\mu}{Tu^2}\right), \\
      & W'_u = 2uV, \\
      & \left.W\right|_{u=0} =\frac{Z}{r_0}, \quad W(1) = V(1) = 0.
    \end{aligned}
  \right.
\end{equation}
From here one have to define a proper value of $\mu$ what can be done by a shooting method.

Zero temperature potential should be calculated separately. Using asymptotics for the Fermi-Dirac functions (See \ref{subsec:Fermi-Dirac}) for the electron density one can obtain:

\begin{multline}
  \label{FTTF_zero_T:eq:density}
  \rho_{TF}(r) = \left.\frac{2}{3}\frac{(2T)^{3/2}}{2\pi^2}\left(\frac{U(r) + \mu}{T}\right)^{3/2}\right|_{T \rightarrow 0} 
  = \\ =
  \frac{2\sqrt2}{3\pi^2}\left(U(r) + \mu \right).
\end{multline}

The boundary problem \eqref{FTTF_finite_T:eq:boundary_problem_reduced2} can be rewritten as follows:
\begin{equation}
  \label{FTTF_zero_T:eq:boundary_problem_reduced}
  \left\{
    \begin{aligned}
      & V'_u = \frac{4a}{3}\left(W + u^2\mu\right)^{3/2}, \\
      & W'_u = 2uV, \\
      & \left.W\right|_{u=0} =\frac{Z}{r_0}, \quad W(1) = V(1) = 0.
    \end{aligned}
  \right.
\end{equation}

\subsection{\label{subsec:FTTF_finite_T_thermodynamic} Thermodynamic functions}

Thermodynamic functions of electrons may be calculated as derivatives of the free energy \eqref{FTTF_finite_T:eq:free_energy} (See \ref{subsec:FTTF_free_energy}). Another way of evaluating them is well described in \cite{Nikiforov:2000}. Here below one can find the final expressions for these quantities:

\begin{equation}
  P_{TF}(V,T) = -F'_V = \frac{(2T)^{5/2}}{6\pi^2}I_{3/2}\left(\frac{\mu(V,T)}{T}\right) 
\end{equation}
\begin{equation}
  P_{TF}(V,0) = \frac{(2\mu)^{5/2}}{15\pi^2}
\end{equation}

\begin{multline}
  \label{FTTF_finite_T:eq:energy}
  E_{TF}(V,T) = \frac{2\sqrt2 VT^{5/2}}{\pi^2}
           \left[
              I_{3/2}\left(\frac{\mu}{T}\right) - \right. \\
           \left. - 3\int_0^1 u^5 I_{3/2} \left(\frac{W + u^2\mu}{u^2 T}\right)du 
           \right] - E_0
\end{multline}
$ E_0 = -0.76874512422 Z^{7/3} $ is supposed to be the energy of an atom at $V \rightarrow \infty$ and $T \rightarrow 0$. 

The integral for the energy depends on the Thomas-Fermi potential. To calculate it with a given precision one can write a differential equation both for the potential and the energy the lower limit with a variable:
\begin{equation}
  \left\{
  \begin{aligned}
    & V'_u = 2au^3T^{3/2}I_{1/2}\left(\frac{W + u^2\mu}{Tu^2}\right), \\
    & W'_u = 2uV, \\
    & E'_u = -3u^5 I_{3/2} \left(\frac{W + u^2\mu}{u^2 T}\right), \\
    & W(1) = V(1) = 0, \\
    & E(1) = \frac{2\sqrt2 VT^{5/2}}{\pi^2}I_{3/2}\left(\frac{\mu}{T}\right) - E_0.
  \end{aligned}
  \right.
\end{equation}
Thus the energy can be obtained as $E_{TF} = E(0)$.

For the energy at $T = 0$ the same procedure should be done:
\begin{multline}
  E_{TF}(V,0) = \frac{4\sqrt2 V}{5\pi^2}
           \left[
              \mu^{5/2} - \right. \\
           \left. - 3\int_0^1 \left(W + u^2\mu\right)^{5/2}du
           \right] - E_0
\end{multline}

\begin{equation}
  \left\{
  \begin{aligned}
    & V'_u = \frac{4a}{3}\left(W + u^2\mu\right)^{3/2}, \\
    & W'_u = 2uV, \\
    & E'_u = -3u^5 \left(W + u^2\mu\right)^{5/2}, \\
    & W(1) = V(1) = 0, \\
    & E(1) = \frac{4\sqrt2 V}{5\pi^2}\mu^{5/2} - E_0.
  \end{aligned}
  \right.
\end{equation}

\section{\label{sec:FTTFQE} Quantum and exchange corrections}

\subsection{\label{subsec:FTTFQE_general} Evaluating corrections to the free energy}

As it was mentioned before, the Thomas-Fermi approximation uses only zero order expansion of a function of two commutators. One could also obtain corrections to the free energy by evaluating higher order terms. Here for simplicity we will consider corrections to the free energy in the form:
\begin{equation}
  \label{FTTFQE_general:eq:shell_free_energy}
  \Delta F = -\int_{-\infty}^{\mu} d\mu' \int \delta \rho(r, \mu')d^3 r.
\end{equation}
where $\delta \rho$ -- correction to the electron density.

To evaluate quantum corrections to the electron density let us consider the density matrix expansion.
With the change of variables $\Phi = (U(r) + \mu)/T$, $y = p^2/(2T) - \Phi$, the result is:
\begin{multline}
  \hat \rho(\hat H)
  =
  \rho(y) + \frac{i\hbar}{2T}\rho''(y)\mathbf p \nabla\Phi + \\ 
   + \frac{\hbar^2}{4T}\rho''(y)\Delta \Phi
   - \frac{\hbar^2}{6T}\rho'''(y)(\nabla \Phi )^2 + \\
   + \frac{\hbar^2}{6T^2}\rho'''(y)(\mathbf p \nabla)^2\Phi 
   - \frac{\hbar^2}{8T^2}\rho^{IV}(y)(\mathbf p \nabla\Phi)^2.
\end{multline}
After the integration \eqref{FTTF_finite_T:eq:denisty} we get:
\begin{multline}
  \rho_e(r) = 
  \frac{\sqrt2T^{3/2}}{\pi^2}
  \left[
    I_{1/2}(\Phi) + \frac{\hbar^2\Delta \Phi}{12T}I''_{1/2}(\Phi) \right. + \\ + \left.\frac{\hbar^2(\nabla\Phi)^2}{24T}I'''_{1/2}(\Phi)
  \right].
\end{multline}

It was shown [Kirzhnits] that the exchange interaction also appears at the order of $\hbar^2$, and one have to evaluate it. 
Thus, the hamiltonian for an electron subsystem must be extended by the exchange interaction:
\begin{equation}
  \label{subsec:FTTFQE:eq:hamiltonian}
  \hat H = \hat H_{TF} + \hat A = \frac{\hat p^2}{2} - U(\mathbf r) + \hat A,
\end{equation}
where the exchange operator can be defined as follows:
\begin{equation}
  \label{subsec:FTTFQE:eq:exchange_operator}
  \hat A(\mathbf r, \hat{\mathbf p})
  =
  \int
  e^{-i\mathbf p \mathbf r/\hbar}
  \frac{4\pi\hbar^2\hat \rho(\hat H')}{|\mathbf p - \mathbf p'|^2}
  e^{i\mathbf p' \mathbf r/\hbar}
  \frac{d^3p'}{(2\pi)^3}.
\end{equation}
The correction to the electron density caused by the exchange interaction can be found as \cite{Nikiforov:2000}:
\begin{multline}
  \label{subsec:FTTFQE:eq:exchange_density}
  \delta \rho_{\mathrm{exc}}(\mathbf r)
  =
  \int
    \left[
      \hat \rho
        \left(\hat H_{TF} + \hat A \right)
      -
      \hat \rho( \hat H_{TF} )
    \right]
    \frac{2d^3p}{(2\pi)^3}
  \simeq \\
  \simeq
  \frac{\partial}{\partial \varepsilon}
  \int \frac{\hat A}{T} \hat \rho (\hat H_{TF}) \frac{d^3p}{(2\pi)^3}
  =
  \frac{2T\hbar^2}{\pi^3}
  \left[I'_{1/2}(\Phi)\right]^2,
\end{multline}

After the integration the correction to density is:
\begin{multline}
  \label{subsec:FTTFQE:eq:density_correction}
  \delta \rho(r) =
  \frac{\sqrt2T^{3/2}}{\pi^2}
  \left[
    \frac{\hbar^2\sqrt2}{\pi\sqrt{T}} \left[ I'_{1/2}\left(\frac{\Phi}{T}\right) \right]^2 + 
  \right. \\
   \left. + \frac{\hbar^2\Delta \Phi}{12T^2}I''_{1/2}\left(\frac{\Phi}{T}\right) + \frac{\hbar^2(\nabla\Phi)^2}{24T^3}I'''_{1/2}\left(\frac{\Phi}{T}\right)
  \right].
\end{multline}

\subsection{\label{subsec:FTTFQE_potential} Corrections to the potential}

The correction to density \eqref{subsec:FTTFQE:eq:density_correction} affects the Thomas-Fermi potential. To keep them self-consistent one have to solve equation for the correction to the potential \cite{Nikiforov:2000}:

\begin{equation}
  \label{TFmodel_correction}
  \left\{
    \begin{aligned}
      \frac{ d^2 \psi }{ dx^2 } &= aT^{1/2}
      \left[
        \psi(x)I_{1/2}'\left(\frac{\phi(x)}{Tx}\right) + \right. \\
        & + \left.T^{1/2}xY_{1/2}'\left(\frac{\phi(x)}{Tx}\right)
      \right] , \\
      \psi(0) &=0, \quad \psi'(1)=\psi(1),
    \end{aligned}
  \right.
\end{equation}

In practice we use the following variables $u^2 = x$, $\psi = Q$, $\psi'_x = Q'_u/(2u) = R$, $\psi''_x = R'_u/(2u)$:

\begin{equation}
  \label{FTTF_Pot:eq:BPreduced2}
  \left\{
    \begin{aligned}
      & R'_u = 2uaT^{1/2}
      \left[
        Q I_{1/2}'\left(\frac{W + u^2\mu}{Tu^2}\right)\right. + \\  
      & + \left.T^{1/2}u^2Y_{1/2}'\left(\frac{W + u^2\mu}{Tu^2}\right)
      \right] , \\
      & Q'_u = 2uR, \\
      & V'_u = 2au^3T^{3/2}I_{1/2}\left(\frac{W + u^2\mu}{Tu^2}\right), \\
      & W'_u = 2uV, \\
      & \left.Q\right|_{u=0} = 0, \\ 
      & Q(1) = R(1) = 0, \\ 
      & W(1) = V(1) = 0.
    \end{aligned}
  \right.
\end{equation}

Asymptotics at $T = 0$:
\begin{equation}
  \label{FTTF_Pot:eq:BPreduced2}
  \left\{
    \begin{aligned}
      & R'_u = 2a
      \left[
        Q \left(W + u^2\mu\right)^{1/2}\right. + \\ + 
        & \left.\frac{22}{3}u\left(W + u^2\mu \right)
      \right] , \\
      & Q'_u = 2uR, \\
      & V'_u = \frac{4a}{3}\left(W + u^2\mu\right)^{3/2}, \\
      & W'_u = 2uV, \\
      & \left.Q\right|_{u=0} = 0, \\
      & Q(1) = R(1) = 0, \\
      & W(1) = V(1) = 0.
    \end{aligned}
  \right.
\end{equation}

\subsection{\label{subsec:FTTFQE_thermodynamics} Corrections to thermodynamic functions}

The expressions for corrections to pressure and energy can be found in \cite{Nikiforov:2000}:
\begin{equation}
  \Delta P(V,T) = \frac{T^{3/2}}{3\pi^3}\left[Q(1)I_{1/2}\left(\frac{\mu}{T}\right) 
           + T^{1/2}Y\left(\frac{\mu}{T}\right)\right]
\end{equation}
\begin{equation}
  \Delta P(V,0) = \frac{\mu^{3/2}}{9\pi^3}\left[2Q(1) + 11\mu^{1/2}\right]
\end{equation}

\begin{multline}
  \label{DE}
  \Delta E(V,T) =
  \frac{VT^{3/2}}{\pi^3}
  \left[
     \int_0^1
     u^3QI_{1/2}
     \left(
       \frac{W + u^2\mu}{Tu^2}
     \right)du
  \right. + \\
  \left.  +
    2T^{1/2}
    \int_0^1
      u^5Y
      \left(
         \frac{W + u^2\mu}{Tu^2}
      \right)du
  \right]
   + \frac{Z\sqrt{2}}{6\pi}R(0)
   - \Delta E_0.
\end{multline}
The integral here is treated similarly to the calculation of the Thomas-Fermi energy \eqref{FTTF_finite_T:eq:energy} by solving the boundary problem with the integration of a thermodynamic function:
\begin{equation}
\label{FTTF_Pot:eq:BPreduced2}
\left\{
  \begin{aligned}
    & R'_u = 2uaT^{1/2}
    \left[
      Q I_{1/2}'\left(\frac{W + u^2\mu}{Tu^2}\right)\right. + \\ 
      & + \left.T^{1/2}u^2Y_{1/2}'\left(\frac{W + u^2\mu}{Tu^2}\right)
    \right] , \\
    & Q'_u = 2uR, \\
    & V'_u = 2au^3T^{3/2}I_{1/2}\left(\frac{W + u^2\mu}{Tu^2}\right), \\
    & W'_u = 2uV, \\
    & \Delta E'_u(V,T,u) = -\frac{VT^{3/2}}{\pi^3}
      \left[
         u^3QI_{1/2}
         \left(
           \frac{W + u^2\mu}{Tu^2}
         \right)
      \right. + \\
    & \left.  +
        2T^{1/2}
          u^5Y
          \left(
             \frac{W + u^2\mu}{Tu^2}
          \right)
      \right], \\
    & Q(1) = R(1) = 0, W(1) = V(1) = 0, \\
    & \Delta E(V,T,1) = \frac{Z\sqrt{2}}{6\pi}R(0) - \Delta E_0.
  \end{aligned}
\right.
\end{equation}

The same procedure is applied to zero temperature asymptotics:
\begin{multline}
  \label{DE}
  \Delta E(V,0) =
  \frac{2V}{3\pi^3}
  \left[
     \int_0^1
     Q
     \left(
       W + u^2\mu
     \right)^{3/2}du
  \right. + \\
  \left.  +
    \int_0^1
      11u
      \left(
         W + u^2\mu
      \right)^2du
  \right]
   + \frac{Z\sqrt{2}}{6\pi}R(0)
   - \Delta E_0.
\end{multline}

\begin{equation}
\label{FTTF_Pot:eq:BPreduced2}
\left\{
  \begin{aligned}
    & R'_u = 2uaT^{1/2}
    \left[
      Q I_{1/2}'\left(\frac{W + u^2\mu}{Tu^2}\right)\right. + \\
      & \left.T^{1/2}u^2Y_{1/2}'\left(\frac{W + u^2\mu}{Tu^2}\right)
    \right] , \\
    & Q'_u = 2uR, \\
    & V'_u = 2au^3T^{3/2}I_{1/2}\left(\frac{W + u^2\mu}{Tu^2}\right), \\
    & W'_u = 2uV, \\
    & \Delta E'_u(V,T,u) = -\frac{2V}{3\pi^3}
      \left[
         Q
         \left(
           W + u^2\mu
         \right)^{3/2}
      \right. + \\
    & \left.  +
          11u
          \left(
             W + u^2\mu
          \right)^2
      \right], \\
    & Q(1) = R(1) = 0, W(1) = V(1) = 0, \\
    & \Delta E(V,T,1) = \frac{Z\sqrt{2}}{6\pi}R(0) - \Delta E_0.
  \end{aligned}
\right.
\end{equation}

\subsection{\label{subsec:self_similarity} Self-similarity and thermal part}

It was found \cite{Kalitkin:IAM:1975} that all thermodynamic functions of the Thomas-Fermi model can be expressed by scaling of the hydrogen ones by atomic number $Z$. The list of these transformations is given below:

\begin{equation}
  \begin{aligned}
  & V_Z = Z^{-1}V_1, & T_Z = Z^{4/3}T_1, \\ 
  & P_Z = Z^{10/3}P_1, & \Delta P_Z = Z^{8/3} \Delta P_1, \\
  & E_Z = Z^{7/3}E_1, & \Delta E_Z = Z^{5/3} \Delta E_1, \\
  & S_Z = Z^{1}S_1,   & \Delta S_Z = Z^{1/3} \Delta S_1, \\
  & \mu_Z = Z^{4/3}\mu_1 & \Delta \mu_Z = Z^{2/3} \Delta \mu_1
  \end{aligned}
\end{equation}

One should notice that during calculations the input data ($V,T$) first is transformed to $Z=1$ values, next all the functions are calculated as for $Z=1$ and then transformed to real value of $Z$. That allow us to use the only one table for precalculated values for $\mu$ for a fast convergence of a shooting method.

The presented approach for an atomic potential and electron density calculation appeared to be good enough for hot and dense matter. But for the matter at normal conditions such description is becoming too rough. In practice cold properties of the model could be removed, and the remaining thermal part can still be useful:

\begin{equation}
  F_T = F(V,T) - F(V,0)
\end{equation}

\section{\label{sec:Shell} Shell structure of a Thomas-Fermi atom}

\subsection{\label{subsec:Shell:thermodynamics} Corrections to thermodynamic functions}
The shell correction to the reduced free energy [] can be found as
\begin{equation}
  \label{Shell:eq:shell_free_energy}
  \Delta F_{sh} = -\int_{-\infty}^{\mu} d\mu' \int \delta \rho_{sh}(r, \mu')d^3 r.
\end{equation}
where the shell correction to density $\delta \rho_{sh}$ also affects the potential.
The boundary problem for corresponding correction to the potential $\delta U$ is:
\begin{equation}
  \label{Shell:eq:BP_potential}
  \left\{
    \begin{aligned} 
      &\Delta \delta U(r) = 4 \pi \delta \rho_t(r), \\
      &\delta U(r_0) = \delta U'(r_0) = 0, \\
      & \left.r\delta U(r)\right|_{r \rightarrow 0} = 0.
    \end{aligned}
  \right.
\end{equation}

The total correction to the electron density $\delta \rho_t$ accounts the influenece of the corrected potential to the Thomas-Fermi density. It is expressed through $\delta \rho_{sh}$ and the first order change in $\rho_{TF}$:
\begin{equation}
  \label{Shell:eq:rho_total}
  \delta \rho_t(r) = \frac{\partial \rho_{TF}(r)}{\partial \mu}(\delta \mu + \delta U(r)) + \delta \rho_{sh}.
\end{equation}
As the number of electrons stays unchanged we have the condition:
\begin{equation}
  \label{Shell:eq:rho_total_norm}
  \int \delta \rho_t(r)d^3r = 0.
\end{equation}

Corrections $\Delta P_{sh}$, $\Delta E_{sh}$ to pressure and energy can be expressed via derivatives \eqref{FTTF_EOS:eq:pressure}, \eqref{FTTF_EOS:eq:energy} of the shell correction to free energy \eqref{Shell:eq:shell_free_energy} (see Appendix):
\begin{equation}
  \label{Shell:eq:P_final}
  \Delta P_{sh} 
  = \rho_{TF}(r_0)\delta \mu_{sh},
\end{equation}
\begin{equation}
  \label{Shell:eq:E_final}
  \Delta E_{sh} 
  = \left[
      \frac32 Z - \int \frac{\partial \rho_{TF}}{\partial \mu}(\mu_{TF} + U(r)) d^3r
    \right]
    \delta\mu_{sh}.
\end{equation}

The correction to the chemical potential $\delta \mu_{sh}$ here is the only unknown property. To evaluate it let us suppose that the chemical potential $\mu$ and the energy spectrum of an atom are already known. Thus the exact number of states of electrons can be written as following:
\begin{equation}
  \label{Shell:eq:number_of_states_exact}
  N(\mu) = 2\sum_{n,l}
        \frac{2l + 1}
        {1 + \exp\left[(\varepsilon_{nl} - \mu)/T)\right]}
\end{equation}
where $ n $, $ l $ -- the principal and the orbital quantum numbers, $\varepsilon_{nl}$ -- the energy levels. If the potential and the electron density are consistent the number of states in a neutral atom equals to the number of electrons:
\begin{equation}
 N(\mu) = Z.
\end{equation}

Owing to the Thomas-Fermi potential is self-consistent the electron density satisfies the condition to the full number of electrons in atomic cell with the appropriate chemical potential $ \mu_{\mathrm{TF}} $:
\begin{equation}
  \label{Shell:eq:number_of_states_TF}
  N_{\mathrm{TF}}(\mu_{\mathrm{TF}}) = \int \rho_{\mathrm{TF}}(r)d^3r = Z
\end{equation}
But if we will take the exact chemical potential $\mu$ this equality breaks.

In order to connect the proper number of states \eqref{Shell:eq:number_of_states_exact} with \eqref{Shell:eq:TF_number_of_states_reduced} one should add the shell correction:
\begin{equation}
 N(\mu) = N_{\mathrm{TF}}(\mu) + \Delta N_{sh}(\mu) = Z
\end{equation}
We suppose here that proper chemical potential slightly differs from the Thomas-Fermi one:
\begin{equation}
  \label{Shell:eq:exact_chemical_potential}
  \mu = \mu_{\mathrm{TF}} + \delta \mu_{sh}.
\end{equation}
The connection between the number of electron states and the shell correction to the chemical potential may be found in the first order expansion \cite{Shpatakovskaya:SovPhysUsp:2012}:
\begin{multline}
    \label{Shell:eq:number_of_states_expansion}
    N(\mu) 
    = N_{\mathrm{TF}}(\mu_{\mathrm{TF}} 
    + \delta \mu_{sh}) 
    + \Delta N_{sh}(\mu_{\mathrm{TF}} 
    + \delta \mu_{sh}) 
    = \\
    = N_{\mathrm{TF}}(\mu_{\mathrm{TF}}) 
    + \frac{\partial N_{\mathrm{TF}}}{\partial \mu}\delta \mu_{sh} 
    + \Delta N_{sh}(\mu_{\mathrm{TF}}).
\end{multline}
Then the equation $ N(\mu) = N_{\mathrm{TF}}(\mu_{\mathrm{TF}}) = Z $ leads us to the following expression for the shell correction to chemical potential:
\begin{equation}
    \label{Shell:eq:chemical_potential_correction}
    \delta \mu_{sh} = -\frac{\Delta N_{sh}(\mu_{\mathrm{TF}})}{\partial N_{\mathrm{TF}}/\partial \mu}.
\end{equation}
In this definition $ \Delta N_{sh}(\mu_{\mathrm{TF}}) $ can be evaluated as:
\begin{equation}
 \label{Shell:eq:number_of_states_correction}
 \Delta N_{sh}(\mu_{\mathrm{TF}}) = N(\mu_{\mathrm{TF}}) - N_{\mathrm{TF}}(\mu_{\mathrm{TF}}).
\end{equation}

\subsection{\label{subsec:Shell:energy_levels} Semiclassical energy levels}

To calculate number of states \eqref{Shell:eq:number_of_states_exact} directly one need to obtain energy levels $\varepsilon_{nl}$. The Bohr-Sommerfeld quantization condition may be used for this task:
\begin{equation}
  \label{alternative:eq:quant_cond}
  S_{nl} = \int_{r_1}^{r_2}p_{nl}(r)dr = \pi\left(n - l - \frac12\right).
\end{equation}
Here is the semiclassical momentum:
\begin{equation}
  \label{alternative:eq:momentum}
  p_{nl}(r) = \sqrt{2\left[\varepsilon_{nl} - U(r) - \frac{(l + 1/2)^2}{2r^2}\right]}.
\end{equation}
By varying $ \varepsilon_{nl} $ for defined $ n $ and $ l $ we can calculate the appropriate value of the action $ S_{nl} $. It is turned out to be strictly monotone function of energy, thus the energy levels can be found from \eqref{alternative:eq:quant_cond} by the bisection method.

In order to guarantee the precision of calculation of the action we developed the following procedure. First we should define the rotate points from the condition $ p_{nl} = 0 $. By replacing $ \lambda = l + 1/2 $ and passing our expression for the potential we get the equation:
\begin{equation}
 \frac{p_{nl}^2}{2} = \varepsilon_{nl} + \frac{W}{u^2} - \frac{\lambda^2}{2r_0^2u^4} = 0
\end{equation}
From here we solve the problem \eqref{FTTF_finite_T:eq:boundary_problem_reduced2} (or \eqref{FTTF_zero_T:eq:boundary_problem_reduced} at $T = 0$) with the already defined $ \mu $ until we find the interval $ [u_2', u_2''] $ where $ p_{nl}^2 $ changes the sign to positive. After that we start from $ u_2'' $ where $ p_{nl}^2~ <~ 0 $ with lower step size and redefine the interval until we get the necessary accuracy $ |u_2' - u_2''|/|u_2' + u_2''| < \varepsilon/2 $. Then the value of the right rotate point supposed to be $ u_2 = u_2' $ in order to have the $ p_{nl}^2 $ from the positive side for further calculations. The same procedure should be done for the next rotate point $ u_1 $.

For the calculation of the action \eqref{alternative:eq:quant_cond} one should store the auxiliary values $W(u_2) = W_2, V(u_2) = V_2 $ and integrate with the Thomas-Fermi potential to the point $ u_1 $. The integral can be transformed to the differential equation:
\begin{equation}
 S_{nl}(u)
 = 2\sqrt{2}r_0\int_u^{u_2}\sqrt{\left(u^2\varepsilon_{nl} 
 + W - \frac{\lambda^2}{2r_0^2u^2}\right)}du,
\end{equation}
and included into the potential problem \eqref{FTTF_finite_T:eq:boundary_problem_reduced2}:
\begin{equation}
  \label{alternative:eq:action_system}
  \left\{
    \begin{aligned}
      & V'_u = 2au^3T^{3/2}I_{1/2}\left(\frac{W + u^2\mu}{Tu^2}\right), \\
      & W'_u = 2uV, \\
      & S'_{nl}(u) = -2\sqrt{2}r_0\sqrt{\left(u^2\varepsilon_{nl} 
 + W - \frac{\lambda^2}{2r_0^2u^2}\right)}, \\
      & W(u_2) = W_2, V(u_2) = V_2, S(u_2) = 0.
    \end{aligned}
  \right.
\end{equation}
The solution $ S_{nl}(u_1) = S_{nl} $ is the desired value of action.
The same problem with zero-temperature potential:
\begin{equation}
  \label{alternative:eq:action_system_Tzero}
  \left\{
    \begin{aligned}
      & V'_u = \frac{4a}{3}\left(W + u^2\mu\right)^{3/2}, \\
      & W'_u = 2uV, \\
      & S'_{nl}(u) = -2\sqrt{2}r_0\sqrt{\left(u^2\varepsilon_{nl} 
 + W - \frac{\lambda^2}{2r_0^2u^2}\right)}, \\
      & W(u_2) = W_2, V(u_2) = V_2, S(u_2) = 0.
    \end{aligned}
  \right.
\end{equation}

\subsection{\label{subsec:Shell:boundary_energy} Boundary energy}

The main difference between between the Thomas-Fermi electron states and the discrete ones appears below some boundary energy, where continuous representation of an electron spectrum fails. Thus, there is no need to calculate all energy levels. Let us suppose that the boundary value of energy which splits the discrete and continuous spectrum is $ \varepsilon_b $. The difference in a number of states \eqref{Shell:eq:number_of_states_correction} then should be considered as:
\begin{equation}
  \Delta N_{sh}(\mu_{\mathrm{TF}}) 
  = \left.N(\mu_{\mathrm{TF}})\right|_{\varepsilon_{nl} < \varepsilon_b} 
  - \left.N_{\mathrm{TF}}(\mu_{\mathrm{TF}})\right|_{\varepsilon < \varepsilon_b}
  \label{Shell:eq:number_of_states_diff_BE}
\end{equation}

\begin{figure}[t]
    \includegraphics[width=1.0\columnwidth]{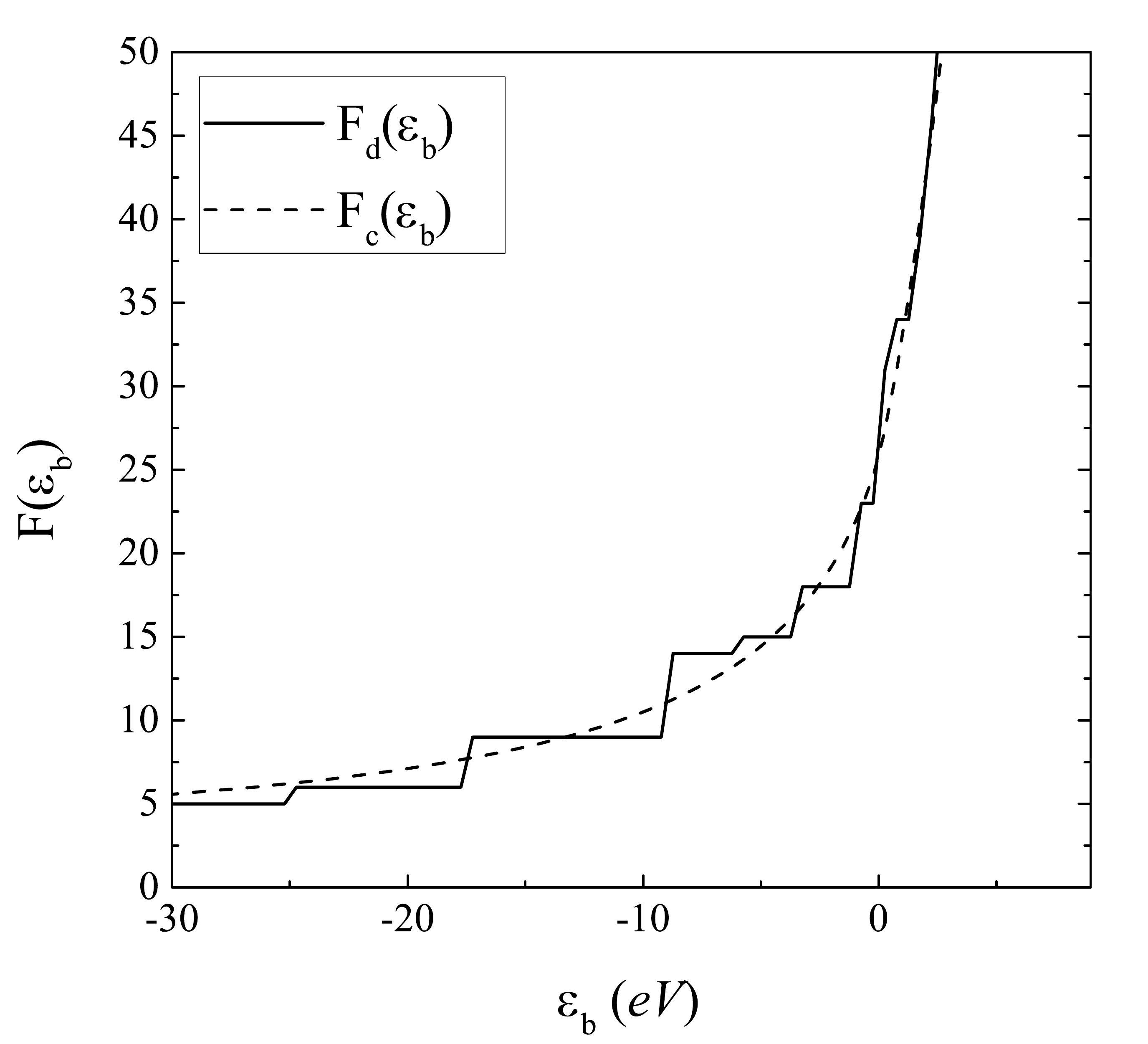}
    \caption{Multiple solutions of equation \eqref{eq:boundary_energy}. Discrete states function: $F_d(\varepsilon_b) = \sum_{n,l}(2l + 1)\theta(\varepsilon_b - \varepsilon_{nl})$, continuous states function: $F_c(\varepsilon_b) =  \iint\theta(\varepsilon_b - \varepsilon)\frac{d^3p}{(2\pi)^3}d^3r $. Calculated for aluminum at $ T = 10\,eV $ and $ \rho = 0.01\rho_0 $.}
    \label{fig:BE_equation}
\end{figure}

\begin{figure}[t]
    \includegraphics[width=1.0\columnwidth]{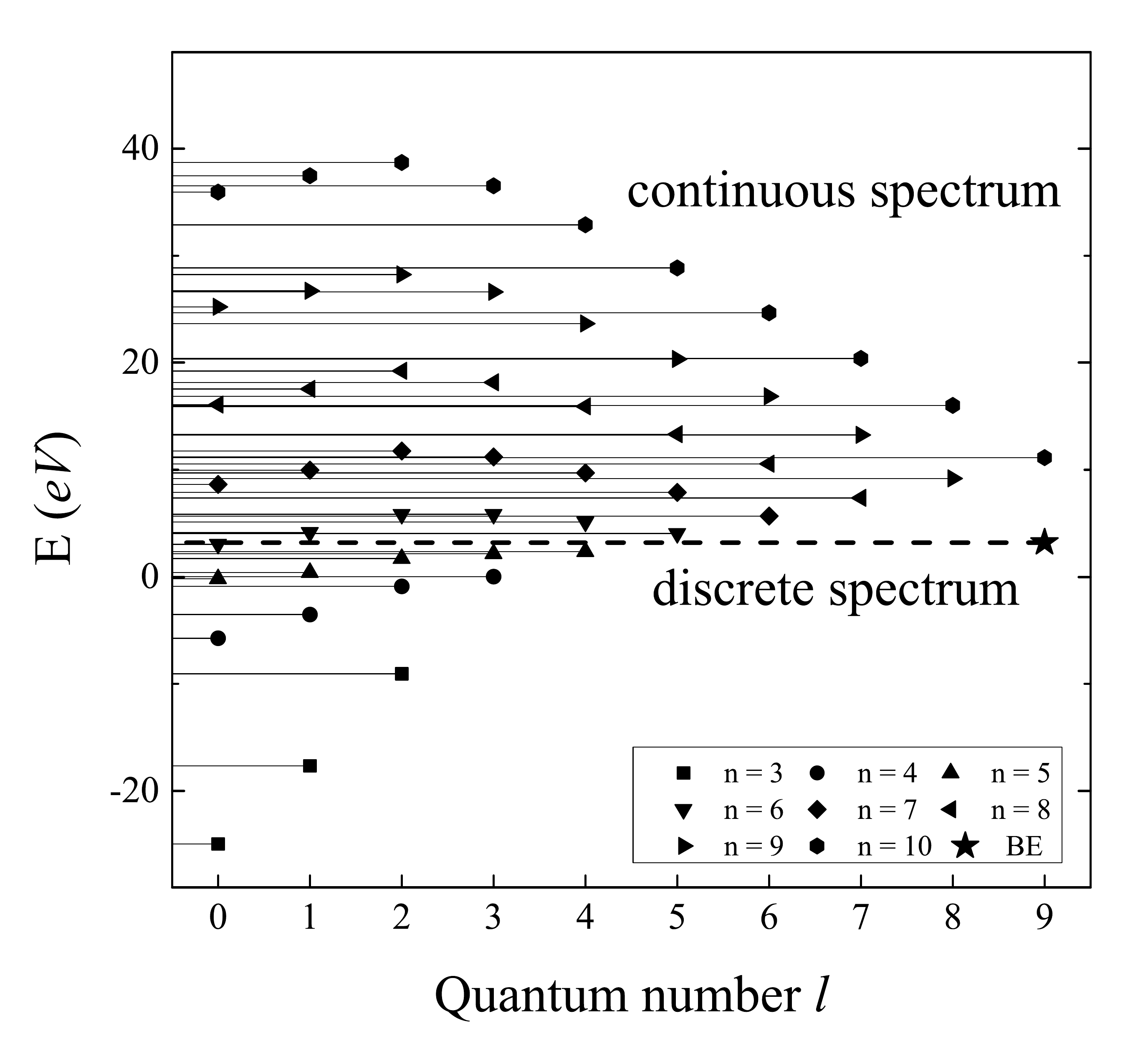}
    \caption{Energy levels in aluminum at $ T = 10\,eV $ and $ \rho = 0.01\rho_0 $ calculated in
    Thomas-Fermi potential. The result of splitting discrete and continuous spectrum is shown 
    as star line.}
    \label{fig:boundary_energy}
\end{figure}

Whether the calculation of exact number of states can be completed with a simple summation in \eqref{Shell:eq:number_of_states_exact}, the Thomas-Fermi electron density from below the boundary energy  is evaluated by replacing the complete Fermi-Dirac integral with the incomplete one:
\begin{multline}
  \left.\rho_{\mathrm{TF}}\right|_{\varepsilon < \varepsilon_b}
  =
  \frac{\sqrt2}{\pi^2}\int_{U(r)}^{\varepsilon_b}\frac{\sqrt{\varepsilon - U(r)}}{1 + \exp{[(\varepsilon - \mu)/T]}}d\varepsilon
  = \\ =
  \frac{\sqrt2 T^{3/2}}{\pi^2}\int_0^{\varepsilon_b/T + W/(Tu^2)}
  \frac{\sqrt{t}dt}{1 + \exp\left[t - \frac{W}{Tu^2} - \frac{\mu}{T}\right]} = \\ 
  = \frac{\sqrt2 T^{3/2}}{\pi^2}I_{1/2}^{inc}\left(\frac{W + u^2\mu}{Tu^2},\frac{W + u^2\varepsilon_b}{Tu^2}\right)
\end{multline}
Similarly, at $ T = 0 $ we get:
\begin{equation}
  \left.\rho_{TF}(u)\right|_{\varepsilon < \varepsilon_b, T = 0} =
  \frac{2\sqrt2}{3\pi^2}\left(\min(\varepsilon_b,\mu) + \frac{W}{u^2}\right)^{3/2}
\end{equation}
Using this bounded density we get for the number of states:
\begin{multline}
  \left.N_{TF}\right|_{\varepsilon < \varepsilon_b} 
  = \int_0^{r_0} 4\pi r^2 \left.\rho_{TF}(r)\right|_{\varepsilon < \varepsilon_b} dr = \\ = 6V \int_0^1 u^5 \left.\rho_{TF}(u)\right|_{\varepsilon < \varepsilon_b} du
  \label{eq:num_TF_states}
\end{multline}
what reduces to the boundary problems:
\begin{equation}
  \label{alternative:eq:action_system}
  \left\{
    \begin{aligned}
      & V'_u = 2au^3T^{3/2}I_{1/2}\left(\frac{W + u^2\mu}{Tu^2}\right), \\
      & W'_u = 2uV, \\
      & N'_{TF}(u) = -\frac{6\sqrt2 VT^{3/2}}{\pi^2}u^5 I_{1/2}^{inc}
        \left(\frac{W + u^2\mu}{Tu^2},\frac{W + u^2\varepsilon_b}{Tu^2}\right), \\
      & W(1) = V(1) = 0, \quad N_{TF}(1) = 0.
    \end{aligned}
  \right.
\end{equation}

\begin{equation}
  \label{alternative:eq:action_system}
  \left\{
    \begin{aligned}
      & V'_u = \frac{4a}{3}\left(W + u^2\mu\right)^{3/2}, \\
      & W'_u = 2uV, \\
      & N'_{TF}(u) = -\frac{4\sqrt2 V}{\pi^2}u^2\left(W + u^2\min(\varepsilon_b,\mu)\right)^{3/2}, \\
      & W(1) = V(1) = 0, \quad N_{TF}(1) = 0.
    \end{aligned}
  \right.
\end{equation}

The proper selection of $ \varepsilon_b $ can be difficult. To handle it accurately the efficient criterion which provides thermodynamical consistency is used here \cite{Nikiforov:2000}. The shell correction $ \Delta N_{sh} $ must stay the same while varying the boundary energy. It appears that the solution $ \varepsilon_b $ of the following equation:
\begin{equation}
  \label{eq:boundary_energy}
    \sum_{n,l}(2l + 1)\theta(\varepsilon_b - \varepsilon_{nl})
    - \iint\theta(\varepsilon_b - \varepsilon)\frac{d^3p}{(2\pi)^3}d^3r = 0
\end{equation}
can guarantee the desired thermodynamic consistency.

The integral here can be reduced to the following:
\begin{equation}
  J = \frac{2\sqrt{2}V}{\pi^2}\int_0^{1} u^2\left(W + u^2\varepsilon_b\right)^{3/2}du
\end{equation}
and we get one more boundary problem:
\begin{equation}
  \label{alternative:eq:action_system}
  \left\{
    \begin{aligned}
      & V'_u = 2au^3T^{3/2}I_{1/2}\left(\frac{W + u^2\mu}{Tu^2}\right), \\
      & W'_u = 2uV, \\
      & J'(u) = -\frac{2\sqrt2 V}{\pi^2}u^2\left(W + u^2\varepsilon_b\right)^{3/2}, \\
      & W(1) = V(1) = 0, \quad J(1) = 0.
    \end{aligned}
  \right.
\end{equation}

\begin{figure}[h]
    \includegraphics[width=0.99\columnwidth]{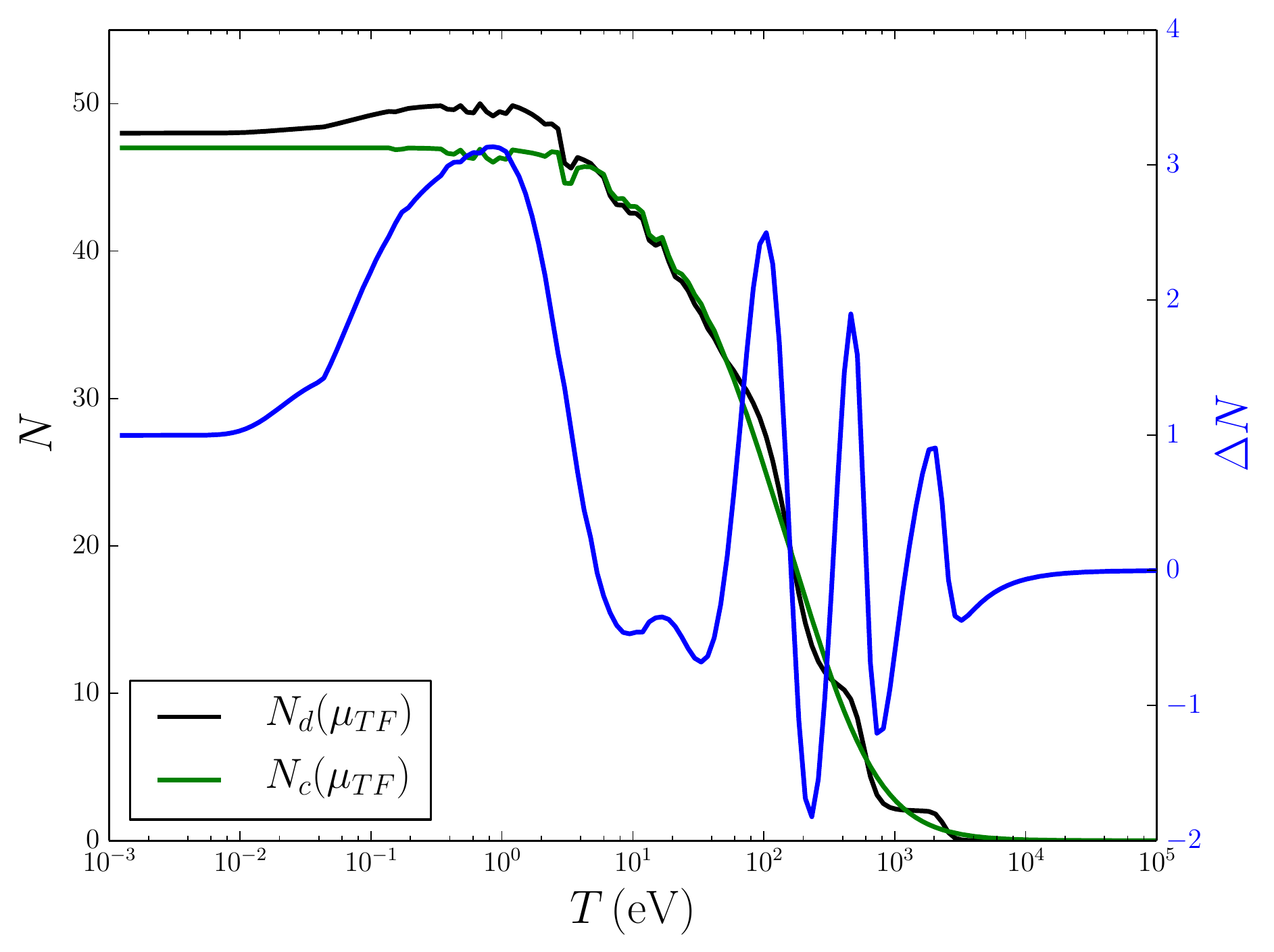}
    \caption{Discrete $N_d$ and continuous $N_c$ numbers of states (left axis) as well as the difference $\Delta N = N_d - N_c$ (right axis) calculated for Ag at $\rho/\rho_0 = 10^{-2}$ with the highest root of Eq.~\eqref{eq:boundary_energy} and $n_{max} = 15$. 
    }
    \label{fig:Ag_Nstates}
\end{figure}

The equation \eqref{eq:boundary_energy} has multiple solutions as it is shown in Fig.~\ref{fig:BE_equation} for Al at $T = 10$\,eV and $\rho/\rho_0 = 10^{-2}$, here $\rho$ is the mass density, $\rho_0$ is the normal density. In practice we select the highest possible root of Eq.~\eqref{eq:boundary_energy} $\varepsilon_b$ in our set of energy levels $\varepsilon_{nl}$ (the maximal value of $n$ in this work is $n_{max} = 15$). In Fig. \ref{fig:Ag_Nstates} one can see that at low temperatures the number of states by Eq.~\eqref{eq:num_TF_states} $N_c=Z$, i.e. all the electron states are placed below $\varepsilon_b$. The number of states by Eq.~\eqref{Shell:eq:number_of_states_exact} $N_d$ here is different from $N_c$ because $N_d$ is calculated at an inconsistent chemical potential $\mu_{TF}$. At some temperature about 1\,eV both $N_d$ and $N_c$ start fluctuating because $\varepsilon_b$ changes from point to point. However, $\Delta N = N_d(\mu_{TF}) - N_c(\mu_{TF})$ remains smooth as the fluctuations of $N_d$ and $N_c$ are correlated due to Eq.~\eqref{eq:boundary_energy}. At very high temperatures $N_d$ and $N_c$ tend to zero because of the excitation of electrons to the states with energies higher than $\varepsilon_b$.

\section{\label{sec:Appendix} Appendix}

\subsection{\label{subsec:Fermi-Dirac} Fermi-Dirac functions}

The Fermi-Dirac function of the order $k > -1$ is defined as:
\begin{equation}
  \label{subsec:eq:Fermi-Dirac}
  I_k(x)=\int_0^\infty\frac{y^kdy}{1+\exp(y-x)}. 
\end{equation}
Following the given definition here is the important relation for the derivatives:
\begin{equation}
  \label{subsec:eq:Fermi-Dirac-derivative}
  I'_k(x)=kI_{k - 1}(x). 
\end{equation}
For the values of $k \leq -1$ the integral diverges. For practical calculations the authors use precise approximations \cite{Antia:1993}. 

One of the difficult cases appears there with the function:
\begin{equation}
  \label{subsec:eq:Y}
  Y(x) = I_{1/2}(x)I'_{1/2}(x) + 6 \int_{-\infty}^x\left[I'_{1/2}(t)\right]^2dt.
\end{equation}
The integral
\begin{equation}
  J(x) = \int_{-\infty}^x\left[I_{-1/2}(t)\right]^2dt
\end{equation}
is evaluated as solution of the differential equation:
\begin{equation}
  \left\{
  \begin{aligned}
   & J'(t) = \left[I_{-1/2}(t)\right]^2, \\
   & J(-a) = 0.
  \end{aligned}
  \right.
\end{equation}
Another problem with the derivative:
\begin{equation}
  \label{subsec:eq:Y_derivative}
  Y'(x) = \frac74I^2_{-1/2}(x) + \frac12I_{1/2}(x)I'_{-1/2}(x).
\end{equation}
The function $I'_{-1/2}(x)$ cannot be evaluated according the general rule \eqref{subsec:eq:Fermi-Dirac-derivative}. In practice the derivative of its rational approximation is calculated.

For calculating the cold part of thermodynamic functions in the Thomas-Fermi theory there are useful approximations for the Fermi-Dirac functions at $x \gg 1$:
\begin{equation}
  I_{-1/2}(x) \sim 2x^{1/2}\left[1 - \frac{\pi^2}{24x^2} - \frac{7\pi^4}{384x^4} + ...\right],
\end{equation}

\begin{equation}
  I_{1/2}(x) \sim \frac{2x^{3/2}}{3}\left[1 + \frac{\pi^2}{8x^2} + \frac{7\pi^4}{640x^4} + ...\right],
\end{equation}

\begin{equation}
  I_{3/2}(x) \sim \frac{2x^{5/2}}{5}\left[1 + \frac{5\pi^2}{8x^2} - \frac{7\pi^4}{384x^4} + ...\right]
\end{equation}

$Y$-function and its derivative can also be evaluated at $x \gg 1$:
\begin{equation}
  Y(x) \sim \frac{11}{3} x^2, \quad Y'(x) \sim \frac{22}{3}x.
\end{equation}

To calculate the number of states below the boundary energy one have to evaluate the incomplete Fermi-Dirac integral:
\begin{equation}
  \label{subsec:eq:Fermi-Dirac-incomplete}
  I^{inc}_{1/2}(x,y)=\int_0^y\frac{z^{1/2}dz}{1+\exp(z-x)}. 
\end{equation}
In our calculations a numerical integration through the differential equation is used as for the function $Y(x)$. 

\subsection{\label{subsec:FTTF_free_energy} The Thomas-Fermi free energy derivatives}

Thermodynamic functions in the Thomas-Fermi model may be calculated as free energy derivatives. First, let us consider pressure:
\begin{multline}
  \label{FTTF_free_energy:eq:pressure}
  -P_{TF} = \frac{\partial F_{TF}}{\partial V} = 
            \frac{\sqrt2}{\pi^2}T^{5/2}
            \left[\Phi(r_0)I_{1/2}(\Phi(r_0))\right. - \\ 
            - \left.\frac23 I_{3/2}(\Phi(r_0)) \right] + 
            \frac{\sqrt2}{\pi^2}T^{5/2}\int d^3r 
            \left[\frac{\partial \Phi}{\partial V}I_{1/2}(\Phi)\right. + \\ +
            \left.\Phi \frac{\partial I_{1/2}(\Phi)}{\partial V}  
            - I_{1/2}(\Phi)\frac{\partial \Phi}{\partial V}
            \right] + \frac{\partial(E_{ee} + E_{ei})}{\partial V} = \\
            = \mu \rho(r_0) - \frac{2\sqrt2}{3\pi^2}T^{5/2}I_{3/2}(\Phi(r_0)) + \\
            \mu \int \frac{\partial \rho}{\partial V}d^3r + 
            \int U(r)\frac{\partial \rho}{\partial V}d^3r 
            + \frac{\partial(E_{ee} + E_{ei})}{\partial V}.
\end{multline}

To evaluate last three summands let us consider helpful expression:
\begin{equation}
  \label{FTTF_free_energy:eq:number_of_states_V_derivative}
  \frac{\partial}{\partial V} \left(\int \rho(r) d^3r = N \right)
  \rightarrow \int \frac{\partial \rho}{\partial V}d^3r = - \rho(r_0).
\end{equation}
After that the first and the third summands from \eqref{FTTF_free_energy:eq:pressure} disappear, and next we have to evaluate the derivative of the potential energy $E_p = E_{ee} + E_{ei} $:
\begin{multline}
  \frac{\partial E_p}{\partial V} 
  = -\frac{1}{2}\frac{\partial}{\partial V}\iint\frac{\rho(r)\rho(r') }{|r' - r|}d^3rd^3r' - \\
    - \frac{\partial}{\partial V} \int \rho(r)U_{i}(r) d^3r = \\
  = -\rho(r_0)U(r_0) - \int U(r)\frac{\partial \rho}{\partial V}d^3r.
\end{multline}
The first summand here is equal to zero, and the second one is opposite to the fourth summand of \eqref{FTTF_free_energy:eq:pressure}. Finally, the expression for the Thomas-Fermi pressure:
\begin{equation}
  \label{FTTF_free_energy:eq:pressure_final}
  P_{TF} = \frac{2\sqrt2}{3\pi^2}T^{5/2}I_{3/2}\left(\frac{\mu + U(r)}{T}\right).
\end{equation}

Next, let us consider the entropy as temperature derivative of the free energy \eqref{FTTF_finite_T:eq:free_energy}:
\begin{multline}
  \label{FTTF_free_energy:eq:entropy}
  -S_{TF} = \frac{\partial F_{TF}}{\partial T} 
         = \frac{5\sqrt2}{2\pi^2}T^{3/2}
         \int
     \left[\Phi(r)I_{1/2}(\Phi(r))\right. - \\ - \left.\frac23 I_{3/2}(\Phi(r))
     \right]d^3r 
   + \frac{\sqrt2}{\pi^2}T^{5/2}\int \Phi(r) \frac{\partial I_{1/2}(\Phi(r))}{\partial T}d^3r + \\
   + \frac{\partial(E_{ee} + E_{ei})}{\partial T}.
\end{multline}
Similarly, as it was done for the pressure \eqref{FTTF_free_energy:eq:number_of_states_V_derivative}, let us consider the derivative:
\begin{equation}
  \label{FTTF_free_energy:eq:number_of_states_T_derivative}
  \frac{\partial}{\partial T} \left(\int \rho(r) d^3r = N \right)
  \rightarrow \int \frac{\partial \rho}{\partial T}d^3r = 0.
\end{equation}
and the potential energy $E_p$ derivative:
\begin{multline}
  \frac{\partial E_p}{\partial T} = 
    -\frac{1}{2}\frac{\partial}{\partial T}\iint\frac{\rho(r)\rho(r') }{|r' - r|}d^3rd^3r' - \\
    - \frac{\partial}{\partial T} \int \rho(r)U_{i}(r) d^3r = -\int U(r)\frac{\partial \rho}{\partial T}d^3r.
\end{multline}
In addition, one have to evaluate the temperature derivative for the second summand in \eqref{FTTF_free_energy:eq:entropy}. To do this it is convenient to use explicit expression for the Thomas-Fermi density \eqref{FTTF_finite_T:eq:density_TF_spheric}:
\begin{equation}
  \frac{\partial \rho}{\partial T} = \frac{3\rho}{2T} 
  + \frac{\sqrt2}{\pi^2}T^{3/2}\frac{\partial I_{1/2}(\Phi(r))}{\partial T},
\end{equation}
from which we get:
\begin{equation}
  \frac{\sqrt2}{\pi^2}T^{5/2}\frac{\partial I_{1/2}(\Phi(r))}{\partial T} 
  = T\frac{\partial \rho}{\partial T} - \frac32 \rho.
\end{equation}
The second and the third summands from \eqref{FTTF_free_energy:eq:entropy} should be reduced as follows:
\begin{multline}
  \int (\mu + U(r))\left(\frac{\partial \rho}{\partial T} 
      - \frac{3\rho}{2T}\right)d^3r 
      - \int U(r)\frac{\partial \rho}{\partial T}d^3r = \\
      = -\frac{3\sqrt2}{2\pi^2}T^{3/2}\int \Phi(r) I_{1/2}(\Phi(r))d^3r.
\end{multline}
Finally, for the Thomas-Fermi entropy we get:
\begin{equation}
  \label{FTTF_free_energy:eq:entropy_final}
  S_{TF} = \frac{\sqrt2}{\pi^2}T^{3/2}
  \int\left[
         \frac{5}{3}I_{3/2}(\Phi(r)) - \Phi(r)I_{1/2}(\Phi(r))
      \right]
      d^3r.
\end{equation}

Now, one can also evaluate the full energy:
\begin{multline}
  \label{FTTF_free_energy:eq:energy}
  E_{TF} = F_{TF} + TS_{TF} = \\
  = \frac{\sqrt2}{\pi^2}T^{5/2}\int I_{3/2}(\Phi(r))d^3r + E_p.
\end{multline}


\subsection{\label{subsec:shell_free_energy} Calculations of shell corrections}

Here we consider the derivation of the expressions for thermodynamic functions:
\begin{equation}
  \Delta P_{sh} = \left.\frac{\partial \Delta F_{sh}}{\partial V}\right|_{T, N}, 
\end{equation}
\begin{equation}
  \Delta E_{sh} = \Delta F_{sh} - T\left.\frac{\partial \Delta F_{sh}}{\partial T}\right|_{V, N}.
\end{equation}
Here we use the shell correction to the free energy \eqref{Shell:eq:shell_free_energy}. For the pressure we have:
\begin{multline}
  \label{shell_free_energy:eq:P1}
  \Delta P_{sh} = \frac{\partial \mu}{\partial V}\int \delta \rho_{sh}(r, \mu)d^3r  
    + \int_{-\infty}^{\mu}d\mu'\int \frac{\partial \delta \rho_{sh}}{\partial V}d^3r + \\
    + \int_{-\infty}^{\mu}\delta \rho_{sh}(r_0, \mu')d\mu'.
\end{multline}
As it was shown in the beginning of \ref{sec:Shell}, $\delta \rho_{sh}$ has explicit dependency of distribution function $ f = (1 + \exp[(E - \mu')/T]) $ and implicit dependency of potential, which is always considered in the expression $(\mu' - U)$, so that:
\begin{multline}
  \frac{\partial}{\partial U}\delta \rho_{sh}\left(\frac{\mu + U}{T}\right) = 
  \frac{\partial}{\partial \mu}\delta \rho_{sh}\left(\frac{\mu + U}{T}\right) = \\
  = \frac{1}{T}\delta \rho'_{sh}\left(\frac{\mu + U}{T}\right),
\end{multline}
\begin{multline}
  \int_{-\infty}^{\mu}\frac{\partial \delta \rho_{sh}}{\partial U}d\mu'  
  = \int_{-\infty}^{\mu}\frac{\partial \delta \rho_{sh}}{\partial \mu}d\mu' = \\
  = \delta \rho_{sh}\left(\frac{\mu + U(r)}{T}\right).
\end{multline}
It allows to transfrom second summand from \eqref{shell_free_energy:eq:P1} because only the potential is a function of volume ($\mu'$ is variable of integration):
\begin{multline}
  \int_{-\infty}^{\mu}d\mu'\int \frac{\partial \delta \rho_{sh}}{\partial U}
  \frac{\partial U}{\partial V}d^3r = 
  \int d^3r \frac{\partial U}{\partial V} 
  \int_{-\infty}^{\mu}\frac{\partial \delta \rho_{sh}}{\partial \mu}d\mu' = \\
  = \int \delta \rho_{sh}(r, \mu)\frac{\partial U}{\partial V} d^3r.
\end{multline}
Next we add the first summand from \eqref{shell_free_energy:eq:P1} to the transformed second:
\begin{equation}
  \Delta P_1 + \Delta P_2 = \int \delta \rho_{sh}(r, \mu) \frac{\partial (\mu + U(r))}{\partial V}d^3r.
\end{equation}
From \eqref{Shell:eq:rho_total} one can express the shell correction to the density and put it into the previous equation:
\begin{multline}
  \Delta P_1 + \Delta P_2 
  = \int \frac{\partial (\mu + U(r))}{\partial V}\left(\delta \rho_t(r)\right. - \\
   - \left.\frac{\partial \rho_{TF}(r)}{\partial \mu}(\delta \mu + \delta U)\right) 
  d^3r.
\end{multline}
The condition \eqref{Shell:eq:rho_total_norm} allows us to get:
\begin{multline}
  \label{shell_free_energy:eq:P2}
  \Delta P_1 + \Delta P_2 
  = \int \delta \rho_t \frac{\partial U}{\partial V} d^3r - \\
  - \delta \mu \int \frac{\partial \rho_{TF}(r)}{\partial \mu} \frac{\partial (\mu + U(r))}{\partial V}d^3r - \\
  - \frac{\partial \mu}{\partial V}\int \frac{\partial \rho_{TF}(r)}{\partial \mu} \delta U d^3r - \\
  - \int \frac{\partial \rho_{TF}(r)}{\partial \mu}\frac{\partial U}{\partial V}\delta U d^3r.
\end{multline}
The derivative of the number of states in Thomas-Fermi model \eqref{Shell:eq:number_of_states_TF}:
\begin{equation}
  \label{shell_free_energy:eq:dNTFdV}
  \frac{\partial}{\partial V}\int \rho_{TF}(r) d^3r =
  \rho_{TF}(r_0) + \int \frac{\partial \rho_{TF}}{\partial \mu}\frac{\partial (\mu + U(r))}{\partial V}d^3r = 0.
\end{equation}
After joining the last two summands from \eqref{shell_free_energy:eq:P2} and using the previous expression we get:
\begin{multline}
  \Delta P_1 + \Delta P_2 
  = \rho_{TF}(r_0)\delta \mu 
  + \int \delta \rho_t \frac{\partial U}{\partial V} d^3r - \\
  - \int \frac{\partial \rho_{TF}}{\partial V}\delta U d^3r.
\end{multline}
The last two summands here may be transformed with poisson equations \eqref{FTTF_finite_T:eq:poisson_general}, \eqref{Shell:eq:BP_potential} and the second Green's formula:
\begin{equation}
  \label{shell_free_energy:eq:Green}
  \int (\varphi \Delta \psi - \psi \Delta \varphi)d^3r 
  = \int_{S} \left(\varphi \frac{\partial \psi}{\partial n} - \psi \frac{\partial \varphi}{\partial n}\right)dS \\
\end{equation}
\begin{equation}
  \frac{1}{4 \pi}\int \left[\frac{\partial U}{\partial V}\Delta \delta U 
  - \delta U\Delta\left(\frac{\partial U}{\partial V}\right)\right]d^3r = 0.
\end{equation}
Finally, we get the shell correction to the pressure:
\begin{equation}
  \Delta P_{sh} = \rho_{TF}(r_0)\delta \mu + \int_{-\infty}^{\mu}\delta \rho_{sh}(r_0, \mu')d\mu'.
\end{equation}
The shell correction to the energy is calculated in similar way. For this we are going to take the partial derivative in temperature:
\begin{equation}
    \label{shell_free_energy:eq:dF_dT}
    \frac{\partial \Delta F_{sh}}{\partial T} = -\frac{\partial \mu}{\partial T}\int d^3r \delta \rho_{sh}(r, \mu) -
    \int_{-\infty}^{\mu}d \mu' \int d^3r \frac{\partial \rho_{sh}(r, \mu')}{\partial T}
\end{equation}
Here we use the same properties of $ \delta \rho_{sh} $ as for the pressure correction:
\begin{equation}
  \label{shell_free_energy:eq:drho_dT}
  \frac{d \delta \rho_{sh}}{d T} = \frac{\partial \delta \rho_{sh}}{\partial T} + 
  \frac{\partial \delta \rho_{sh}}{\partial U}\frac{\partial U}{\partial T} = 
  \frac{\partial \delta \rho_{sh}}{\partial T} - 
  \frac{\partial \delta \rho_{sh}}{\partial \mu'}\frac{\partial U}{\partial T}
\end{equation}
and pass it to the \eqref{shell_free_energy:eq:dF_dT}: 
\begin{multline}
  \label{shell_free_energy:eq:dF_dT2}
  \frac{\partial \Delta F_{sh}}{\partial T} = -\frac{\partial \mu}{\partial T}\int d^3r \delta \rho_{sh}(r, \mu) - \\ -
  \int_{-\infty}^{\mu}d \mu' \int d^3r 
  \left(
    \frac{\partial \delta \rho_{sh}}{\partial T} - 
    \frac{\partial \delta \rho_{sh}}{\partial \mu'}\frac{\partial U}{\partial T}
  \right) = \\ =
  -\int d^3r \delta \rho_{sh}(r,\mu)\frac{\partial (\mu - U)}{\partial T} - \\ -
  \int_{-\infty}^{\mu}d \mu' \int d^3r \frac{\partial \delta \rho_{sh}}{\partial T}.
\end{multline}
The Poisson formula can be used for other thermodynamic properties in order to obtain explicit dependencies. For the energy and density corrections we have:
\begin{equation}
  \label{shell_free_energy:eq:rho_shell}
  \delta \rho_{sh}(r) = \frac{1}{2\pi}\sum'_{k,s}\int_{-\infty}^{+\infty}\frac{J_{ks}(r, \varepsilon)d\varepsilon}
                        {1 + \exp[(\varepsilon - \mu)/T]},
\end{equation}
\begin{equation}
  \label{shell_free_energy:eq:E_shell}
  \Delta E_{sh} = \frac{1}{2\pi}\int d^3r \int_{-\infty}^{+\infty}\sum'_{k, s}
                  \frac{\varepsilon J_{ks}(r, \varepsilon)d\varepsilon}{1 + \exp[(\varepsilon - \mu)/T]},
\end{equation}
where the summands are:
\begin{equation}
  J_{ks}(r, \varepsilon) = \int (2l + 1)|R_{\varepsilon l}(r)|^2 dl.
\end{equation}
Let us explicitly differentiate $\delta \rho_{sh}$ in the end of \eqref{shell_free_energy:eq:dF_dT2} by taking into account the following:
\begin{equation}
  \frac{\partial f}{\partial T} = \frac{\partial f}{\partial \mu'}\frac{\varepsilon - \mu'}{T}.
\end{equation}
After the integration over $\mu'$ the expression looks like:
\begin{multline}
  \label{shell_free_energy:eq:dF_dT3}
  \frac{\partial \Delta F_{sh}}{\partial T} 
  = -\int d^3r \delta \rho_{sh}(r,\mu)\frac{\partial (\mu - U)}{\partial T} - \frac{\Delta E_{sh}}{T} + \\
    + \frac{\mu}{T}\int d^3r \delta \rho_{sh} (r, \mu) + \frac{\Delta F_{sh}}{T}
\end{multline}
The next step is to express the $\delta \rho_{sh}$ from \eqref{Shell:eq:rho_total} and calculate the first summand in previous formula:
\begin{multline}
    \label{shell_free_energy:eq:expr}
    -\int d^3r \delta \rho_{sh}(r,\mu)\frac{\partial (\mu - U)}{\partial T} = \\ =
    \frac{\partial \mu}{\partial T} \int d^3 r \frac{\partial \rho_{TF}}{\partial \mu}
    (\delta \mu - \delta U) + \\
    + \int d^3r 
    \left[
      \delta \rho_{t} - \frac{\partial \rho_{TF}}{\partial \mu}(\delta \mu - \delta U)
    \right]
    \frac{\partial U}{\partial T} = \\ =
    \int d^3 r \frac{\partial \rho_{TF}}{\partial \mu}
    (\delta \mu - \delta U)\frac{\partial (\mu - U)}{\partial T} + \\
    + \int d^3 r \delta \rho_t \frac{\partial U}{\partial T}
\end{multline}
The last summand here can be transformed with a help of the second Green's function \eqref{shell_free_energy:eq:Green} and Poisson equation
\eqref{Shell:eq:BP_potential}:
\begin{equation}
  \int \delta \rho_{sh} \frac{\partial U}{\partial T} d^3r = \int \delta U \frac{d \rho_{TF}}{dT}d^3 r
\end{equation}
For the Thomas-Fermi density we have the following expression:
\begin{equation}
  \frac{d \rho_{TF}}{d T} = \frac{\partial \rho_{TF}}{\partial T} 
  + \frac{\partial \rho_{TF}}{\partial \mu}\frac{\partial(\mu - U)}{\partial T},
\end{equation}
and derivative of the normalization condition \eqref{Shell:eq:number_of_states_TF}:
\begin{equation}
  \int \frac{\partial \rho_{TF}}{\partial \mu} d^3r 
  = -\int \frac{\partial \rho_{TF}}{\partial \mu}\frac{\partial(\mu - U)}{\partial T} d^3 r.
\end{equation}
Let us put all these expressions together in \eqref{shell_free_energy:eq:expr}:
\begin{multline}
  \int d^3 r \frac{\partial \rho_{TF}}{\partial \mu}
    (\delta \mu - \delta U)\frac{\partial (\mu - U)}{\partial T} + \\ +
    \int \delta U \left[\frac{\partial \rho_{TF}}{\partial T} + 
        \frac{\partial \rho_{TF}}{\partial \mu}\frac{\partial (\mu - U)}{\partial T}\right]
        d^3r = \\ =
        \delta \mu \int \frac{\partial \rho_{TF}}{\partial \mu}\frac{\partial (\mu - U)}{\partial T}d^3r + \\
        + \int \delta U \frac{\partial \rho_{TF}}{\partial T}d^3r = -\int (\delta \mu - \delta U) \frac{\partial \rho_{TF}}{\partial T}d^3r
\end{multline}
The derivative of Thomas-Fermi density can be calculated explicitly:
\begin{equation}
  \frac{\partial \rho_{TF}}{\partial T} = \frac32\frac{\rho_{TF}}{T} - \frac{\partial \rho_{TF}}{\partial \mu}\frac{(\mu - U)}{T},
\end{equation}
and for the first summand in \eqref{shell_free_energy:eq:dF_dT3} we have
\begin{equation}
  -\int (\delta \mu - \delta U)\left[\frac32\frac{\rho_{TF}}{T} - \frac{\partial \rho_{TF}}{\partial \mu}\frac{(\mu - U)}{T}\right].
\end{equation}
The third summand may be transformed with a help of the condition \eqref{Shell:eq:rho_total_norm}:
\begin{equation}
  \frac{\mu}{T}\int d^3r \delta \rho_{sh} (r, \mu) = 
  -\frac{\mu}{T}\int \frac{\partial \rho_{TF}}{\partial \mu}(\delta \mu - \delta U)d^3r
\end{equation}
At last we get the following expression for the derivative of the free energy correction:
\begin{multline}
  \frac{\partial \Delta F_{sh}}{\partial T} = -\frac{1}{T}\int (\delta \mu - \delta U)
  \left[\frac32 \rho_{TF} + \frac{\partial \rho_{TF}}{\partial \mu}U\right]d^3r + \\
  + \frac{\Delta F_{sh} - \Delta E_{sh}}{T},
\end{multline}
and for the total correction to energy:
\begin{equation}
  \Delta E = \int (\delta \mu - \delta U)\left[\frac32\rho_{TF} + \frac{\partial \rho_{TF}}{\partial \mu}U\right]d^3r + \Delta E_{sh},
\end{equation}
or slightly transformed with $U = -(\mu - U) + \mu$:
\begin{multline}
  \label{shell_free_energy:eq:dE_full}
  \Delta E = \int (\delta \mu - \delta U)\left[\frac32\rho_{TF} - \frac{\partial \rho_{TF}}{\partial \mu}(\mu - U)\right]d^3r + \\
  + \Delta E_{sh} - \mu \Delta N_{sh}.
\end{multline}
The difference of last two summands and $\delta U$ are considered to be small enough and was not accouned in \eqref{Shell:eq:E_final}.

%

\end{document}